\documentclass[12pt]{iopart}

\usepackage{natbib}
\usepackage{graphicx}
\usepackage{iopams}

\begin{document}

\title{Reciprocal locomotion of dense swimmers in Stokes flow}

\author{David Gonzalez-Rodriguez$^1$ and Eric Lauga$^2$}

\address{$^1$ Department of Civil and Environmental Engineering, Massachusetts Institute of Technology, Cambridge, MA 02139, USA}
\address{$^2$ Department of Mechanical and Aerospace Engineering, University of California, San Diego, 9500 Gilman Dr., La Jolla, CA 92093-0411, USA}
\ead{\mailto{davidgr@alum.mit.edu}, \mailto{elauga@ucsd.edu}}
\begin{abstract}
Due to the kinematic reversibility of Stokes flow, a body executing a reciprocal motion (a motion in which the sequence of body configurations remains identical under time reversal) cannot propel itself in a viscous fluid in the limit of negligible inertia; this result is known as Purcell's scallop theorem. In this limit, the Reynolds numbers based on the fluid inertia and on the body inertia are all zero. Previous studies characterized the breakdown of the scallop theorem with fluid inertia. In this paper we show that, even in the absence of fluid inertia, certain dense bodies undergoing reciprocal motion are able to swim. Using Lorentz's reciprocal theorem, we first derive the general differential equations that govern the locomotion kinematics of a dense swimmer. We demonstrate that no reciprocal swimming is possible if the body motion consists only of tangential surface deformation (squirming). We then apply our general formulation to compute the locomotion of four simple swimmers, each with a different spatial asymmetry, that perform normal surface deformations. We show that the resulting swimming speeds (or rotation rates) scale as the first power of a properly defined ``swimmer Reynolds number'', demonstrating thereby a continuous breakdown of the scallop theorem with body inertia.
\end{abstract}

\maketitle

\section{Introduction\label{sec1}}

At the small scales of swimming microorganisms, the inertial mechanisms of locomotion used by larger animals become ineffective. Fluid flow in the limit of zero Reynolds number is governed by the Stokes equations, which are both linear and independent of time. In this limit, Newton's equations of body motion reduce to balances of forces and torques, which depend on time only through the sequence of body configurations. Due to these properties, a periodic motion in which the sequence of body configurations remains identical under time reversal (termed a ``reciprocal'' motion) yields a net translation equal to its opposite, and therefore equal to zero. This result is known as Purcell's ``scallop theorem'' \cite{Purcell77}. In particular, any periodic motion of an organism with one degree of freedom, such as a scallop, is reciprocal. Such an organism cannot propel itself in the limit of zero Reynolds number.

Consider a swimmer (i.e., a self-propelled, deforming body) of density  $\rho_{\rm p}$ and characteristic size $a$ immersed in a fluid of density $\rho \ll \rho_{\rm p}$ and dynamic viscosity $\mu$. The swimmer executes a reciprocal motion of amplitude $A$ and radian frequency $\omega$. There are three relevant Reynolds numbers that characterize this oscillatory motion \cite{Lauga07}. The ``unsteady'' Reynolds number, $Re_{\rm \omega}\equiv \rho a^2 \omega / \mu$, is the scale ratio of the unsteady terms to the viscous terms in the Navier-Stokes equations. The ``advective'' (or ``frequency'' \cite{Lauga07}) Reynolds number, $Re_{\rm f}\equiv \rho a A \omega / \mu$, is the scale ratio of the nonlinear advective terms to the viscous terms. Finally, the ``particle'' Reynolds number (or ``Stokes number'' \cite{KochHill01}), $Re_{\rm p} \equiv \rho_{\rm p} a^2 \omega / \mu$, is the scale ratio of the particle inertia to the viscous forces on the particle.

How much inertial force is necessary for a  reciprocal motion to become propulsive? Childress and Dudley \cite{ChildressDudley04} addressed this question for the case of a \emph{symmetric}, reciprocal flapper and concluded that the breakdown of the scallop theorem occurs above a finite threshold of the ``advective'' Reynolds number, $Re_{\rm f}$, of order unity. As discussed by Childress and Dudley, the breakdown of the scallop theorem is relevant to the study of swimming organisms that cross the $Re_{\rm f}$ threshold as they grow, such as the larvae of certain molluscs \cite{ChildressDudley04}, crustaceans \cite{Williams94}, tunicats \cite{McHenryAziziEtAl03}, or fish \cite{Hunter72,Weihs80}. The existence of a finite threshold above which the scallop theorem becomes invalid has been confirmed in laboratory experiments \cite{VandenbergheZhangEtAl04,VandenbergheChildressEtAl06} and numerical simulations \cite{AlbenShelley05}. In contrast, Lauga \cite{Lauga07} devised examples of oscillatory reciprocal motions \emph{with broken spatial symmetries} for which the net translational velocity is proportional to $Re_{\rm f}^\alpha$, for a certain $\alpha>0$, thus demonstrating a continuous breakdown of the scallop theorem for any $Re_{\rm f}>0$ (see also \cite{Roper07}). In this paper, we study the breakdown of the scallop theorem with $Re_{\rm p}$ in the presence of spatial asymmetries. While Lauga \cite{Lauga07} studied the limit of $\{Re_{\rm \omega}, Re_{\rm p}\} \ll Re_{\rm f} \ll 1$, here we consider dense swimmers ($\rho_{\rm p} / \rho \gg 1$ and $\rho_{\rm p} / \rho \gg A/a$) for which $\{Re_{\rm \omega}, Re_{\rm f}\} \ll Re_{\rm p} \ll 1$. We consider therefore the limit of negligible fluid inertia and show that locomotion is possible nonetheless. As in previous studies of the dynamics of simple swimmers \cite[e.g.,][]{NajafiGolestanian04,AvronKennethEtAl05,DreyfusBaudryEtAl05}, the examples presented here do not necessarily resemble real organisms, but they are intended to illustrate the theory and provide physical insight. Analogous to \cite{Lauga07}, our examples show a breakdown of the scallop theorem for any arbitrarily small $Re_{\rm p}>0$, since the net translational velocity (or rotation rate) for all our examples is proportional to $Re_{\rm p}$.

In section \ref{sec2}, we derive a general framework, using Lorentz's reciprocal theorem \cite{HappelBrenner65, StoneSamuel96}, to describe the motion of a dense swimmer. In section \ref{sec3}, we apply our general framework to study four examples of spatially asymmetric, dense reciprocal swimmers and show that they experience directed motion for arbitrarily small values of $Re_{\rm p}$. The nonlinear interaction mechanism responsible for locomotion in the four examples is summarized in section \ref{sec4}.

\section{Reciprocal theorem for a dense swimmer\label{sec2}}

Consider the limit of a dense swimmer, $\{Re_{\rm \omega}, Re_{\rm f}\} \ll Re_{\rm p} \ll 1$, introduced in section \ref{sec1}. If the fluid inertia is neglected, Lorentz's reciprocal theorem \cite{HappelBrenner65} can be applied to relate the instantaneous dynamics of the swimmer to those of a towed, rigid body of the same shape \cite{StoneSamuel96}. Let the rigid body be towed with a constant force, $\boldsymbol{\hat F}$, plus a constant torque about a point $P$ in the body, $\boldsymbol{\hat L_P}$. Assume the swimmer and the rigid body to be immersed in fluids of the same kinematic viscosity. The surface velocities and stresses on the swimmer, $(\bi{u},\bsigma)$, are related to those on the rigid body, $(\boldsymbol{\hat{u}},\boldsymbol{\hat{\sigma}})$, by
\begin{equation}
\int_{S(t)} \bi{n} \cdot \boldsymbol{\hat{\sigma}} \cdot \bi{u} \; \rmd S = \int_{S(t)} \bi{n} \cdot \bsigma \cdot \boldsymbol{\hat{u}} \; \rmd S, \label{eq2.1}
\end{equation}
where $S(t)$ is the instantaneous location of the swimmer surface and $\bi{n}$ is the normal vector to it. The surface velocity of the swimmer can be written as $\bi{u}=\boldsymbol{U_{\rm P}}+\boldsymbol{\it \Omega}\times\left(\boldsymbol{r}-\boldsymbol{r_{\rm P}}\right)+\boldsymbol{u'}$, where $\boldsymbol{U_{\rm P}}$ is the translational velocity of the reference point $P$, $\boldsymbol{\it \Omega}$ is the angular velocity, $\boldsymbol{r}$ is the position vector with an origin fixed at the inertial frame, and $\boldsymbol{u'}$ is the deformational component of the surface velocity. Note that the values of $\boldsymbol{u'}$ depend on the choice of $P$. All time derivatives are referred to the inertial frame. The surface velocity of the towed, rigid body is $\boldsymbol{\hat u}=\boldsymbol{\hat U_{\rm P}} + \boldsymbol{\it \hat \Omega}\times\left(\boldsymbol{r}-\boldsymbol{r_{\rm P}}\right)$. For a dense swimmer of constant mass, $m$, and homogeneous density, balances of forces and torques require
\numparts
\begin{equation}
\int_{S(t)} \bi{n} \cdot \boldsymbol{\sigma} \; \rmd S = m \frac{\rmd \boldsymbol{u_{\rm G}}}{\rmd t} \label{eq2.2}
\end{equation}
\begin{equation}
\int_{S(t)} \left(\boldsymbol{r}-\boldsymbol{r_{\rm P}}\right) \times \left(\bi{n} \cdot \boldsymbol{\sigma}\right) \; \rmd S = \left(\boldsymbol{r_{\rm G}}-\boldsymbol{r_{\rm P}}\right) \times \left(m \frac{\rmd \boldsymbol{u_{\rm G}}}{\rmd t}\right) + \frac{\rmd \left( \boldsymbol{I_{\rm G}}\cdot \boldsymbol{\it \Omega}\right)}{\rmd t}, \label{eq2.2.5}
\end{equation}
\endnumparts
where $\boldsymbol{r_{\rm G}}$ and $\boldsymbol{u_{\rm G}}$ are the position vector and velocity of the swimmer's centre of mass, $G$, and $\boldsymbol{I_{\rm G}}$ is the inertia tensor referred to $G$. Introducing (\ref{eq2.2}) and (\ref{eq2.2.5}) into (\ref{eq2.1}) results in
\begin{eqnarray}
\fl \left(m \frac{\rmd \boldsymbol{u_{\rm G}} }{\rmd t}\right) \cdot \left(\boldsymbol{\hat{U}_{\rm P}} + \boldsymbol{\it \hat \Omega} \times \left(\boldsymbol{r_{\rm G}}-\boldsymbol{r_{\rm P}}\right)\right)+\frac{\rmd \left( \boldsymbol{I_{\rm G}} \cdot \boldsymbol{\it \Omega}\right)}{\rmd t}\cdot \boldsymbol{\it \hat \Omega}  - \boldsymbol{\hat F} \cdot \boldsymbol{U_{\rm P}} - \boldsymbol{\hat L_{\rm P}} \cdot \boldsymbol{\it \Omega} \nonumber \\ = \int_{S(t)} \bi{n} \cdot \boldsymbol{\hat{\sigma}} \cdot \boldsymbol{u'} \; \rmd S. \label{eq2.3}
\end{eqnarray}
This derivation is valid in the absence of a body force. In the presence of a homogeneous gravitational acceleration, $\boldsymbol{g}$, the terms $m \, \rmd \boldsymbol{u_{\rm G}} / \rmd t$ in (\ref{eq2.2}), (\ref{eq2.2.5}), and (\ref{eq2.3}) should be replaced by $m \, (\rmd \boldsymbol{u_{\rm G}} / \rmd t - \boldsymbol{g})$. In (\ref{eq2.3}), the rigid body velocities, $\boldsymbol{\hat U_{\rm P}}$ and $\boldsymbol{\it \hat \Omega}$, are arbitrary. By alternately taking $\boldsymbol{\it \hat \Omega}=\boldsymbol{0}$ and $\boldsymbol{\hat U_{\rm P}}=\boldsymbol{0}$, (\ref{eq2.3}) yields two differential equations for the translational and rotational velocities of the swimmer, $\boldsymbol{U_{\rm P}}$ and $\boldsymbol{\it \Omega}$.

As an example, consider the application of (\ref{eq2.3}) to a squirming sphere of radius $a$, which undergoes purely tangential deformation. Take $P\equiv G$, which is assumed to remain at the sphere centre. The differential equation for $\boldsymbol{U_{\rm G}}$ is obtained by choosing $\boldsymbol{\it \hat \Omega}=\boldsymbol{0}$, for which $\boldsymbol{\hat L_{\rm P}}=\boldsymbol{0}$. Then, $\boldsymbol{\hat F}=-6 \pi \mu a \boldsymbol{\hat U_{\rm G}}$ and $\boldsymbol{n}\cdot \boldsymbol{\hat \sigma} \cdot \boldsymbol{\hat U} = -3\mu / (2 a)$, which yields
\begin{equation}\frac{2a^2 \rho_{\rm p}}{9\mu} \frac{\rmd \boldsymbol{U_{\rm G}}}{\rmd t} + \boldsymbol{U_{\rm G}} = -\frac{1}{4\pi a^2}\int_{S_0} \boldsymbol{u'} dS, \label{eq2.4}
\end{equation}
where $S_0$ represents the sphere surface. The differential equation for $\boldsymbol{\it \Omega}$ is obtained by choosing $\boldsymbol{\hat U_{\rm G}}=\boldsymbol{0}$, for which $\boldsymbol{\hat F}=\boldsymbol{0}$. Then, $\boldsymbol{\hat L_{\rm G}}=-8 \pi \mu a^3 \boldsymbol{\it \hat \Omega}$ and $\boldsymbol{n}\cdot \boldsymbol{\hat \sigma} \cdot (\boldsymbol{\it \hat \Omega} \times \boldsymbol{n}) = -3\mu$, which yields
\begin{equation}\frac{a^2 \rho_{\rm p}}{15\mu} \frac{\rmd \boldsymbol{\it \Omega}}{\rmd t} + \boldsymbol{\it \Omega} = -\frac{3}{8\pi a^3}\int_{S_0} \boldsymbol{n} \times \boldsymbol{u'} dS. \label{eq2.5}
\end{equation}
Note that (\ref{eq2.4}) and (\ref{eq2.5}) reduce to equations 4 and 6 of \cite{StoneSamuel96} in the limit of $Re_p = 0$. Note also that, for a reciprocal squirming deformation, the time averages of (\ref{eq2.4}) and (\ref{eq2.5}) yield zero net translation and rotation. The same conclusion is obtained from (\ref{eq2.3}) for any reciprocal squirmer, since $S(t)=S_0$ is constant, and $\boldsymbol{\hat F}$, $\boldsymbol{\hat L_{\rm p}}$, and $\boldsymbol{\hat \sigma}$ remain constant over the period of motion for constant $\boldsymbol{\hat U_{\rm P}}$ and $\boldsymbol{\hat \Omega}$. Therefore, a dense reciprocal swimmer needs to undergo deformation normal to its surface in order to propel itself.

\section{Examples of dense reciprocal swimmers\label{sec3}}

The general framework derived in the previous section is here applied to four specific swimmers: two unequal spheres moving along a straight line, two unequal spheres moving along a circumference, a scallop-like swimmer, and a deforming sphere. These simple examples illustrate reciprocal locomotion in the absence of fluid inertia; the goal is therefore to compute the net translational velocity of each swimmer as a function of $Re_{\rm p}$. For simplicity, gravity is not accounted for in these examples; the inclusion of gravity is straightforward, and its effect is discussed at the end of section \ref{sec4}.

\subsection{Translation of two unequal spheres\label{sec3.1}}

\begin{figure} [htbp]
\begin{center}
\includegraphics[width=1.5in]{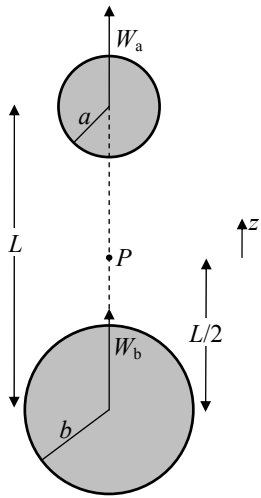}
\end{center}
\caption{Geometry of the two unequal spheres moving along their line of centres ($z$-direction).} \label{fig1}
\end{figure}

Consider the swimmer shown in figure \ref{fig1}, which consists of two unequal spheres. The spheres have radii $a$ and
$b$, which are different but of comparable magnitude, that is,
$\beta \equiv b/a = \Or(1)$. Note $\beta\neq 1$ is required to break the spatial symmetry and yield net motion. The time-dependent distance
between the sphere centres, $L(t)$, is large compared with the
sphere radii, so that $L=\Or(a/\epsilon)$, where $\epsilon \equiv a/\langle L \rangle \ll 1$, and the angular brackets $\langle \rangle$ denote a time average over the period of motion. The two spheres are able to exert equal and opposite forces on each other, so that $L(t)$ is a prescribed periodic function of time with radian frequency $\omega$. The spheres
move with velocities $W_{\rm a}(t) = W(t)+w'(t)$ and $W_{\rm b} = W(t)-w'(t)$,
where $w' = 1/2 \; dL/dt$, and $W(t)$ is the unknown translational
velocity of the reference point, $P$, which is chosen at the midpoint between the sphere centres. The parameter of interest is the net translational velocity, $\langle W(t) \rangle$.

 The reciprocal theorem (\ref{eq2.3}) relates the dynamics of interest to those of a system of two spheres separated a distance $L$ being towed at a constant speed $\hat{W}$ along their line of centres. By computing the effect of each sphere on the
other as that of an equivalent point force, the hydrodynamic resistance forces in $z$-direction on the rigid, towed spheres are \cite{HappelBrenner65}
\numparts
\begin{eqnarray}
\hat F^{\rm h}_{\rm a} &=& 6 \pi \mu a \hat{W} \left[-1 + \frac{3 b}{2 L} + O\left(\epsilon^2\right)\right]  \label{eq3.1} \\
\hat F^{\rm h}_{\rm b} &=& 6 \pi \mu b \hat{W} \left[1 - \frac{3 a}{2 L} + O\left(\epsilon^2\right)\right] \label{eq3.2}.
\end{eqnarray}
\endnumparts

Application of (\ref{eq2.3}) to this swimmer yields
\begin{eqnarray}
\fl \rho_{\rm p}\frac{4}{3}\left[\left(a^3+b^3\right)\frac{\rmd W}{\rmd t}+\left(a^3-b^3\right)\frac{\rmd w'}{\rmd t}\right]\hat{W} - \left(\hat F^{\rm h}_{\rm a}+\hat F^{\rm h}_{\rm b}\right)W \nonumber \\ = \left(\hat F^{\rm h}_{\rm a}-\hat F^{\rm h}_{\rm b}\right)w'. \label{eq3.3}
\end{eqnarray}
Next, we non-dimensionalize (\ref{eq3.3}) using $a$ and $1/\omega$
as the length and time scales. The non-dimensional variables and
parameters are $\tau \equiv \omega t$, $\beta \equiv b/a$, $\tilde{W} \equiv \epsilon W/ (a \omega)$,
and $\lambda(\tau) = \epsilon L / a$. We expand $\tilde{W}$ in powers of $Re_{\rm p}\equiv \rho_{\rm p}
a^2 \omega / \mu \ll \epsilon \ll 1$, i.e.,
$\tilde{W} =\tilde{W}_0 + Re_{\rm p} \tilde{W}_1 + \ldots$. With
this, (\ref{eq3.3}) becomes
\begin{eqnarray}
\fl Re_{\rm p} \lambda (1+\beta^3)\left(\frac{\rmd \tilde{W}_0}{\rmd \tau}+\ldots\right) +
\frac{9}{2}[(1+\beta)\lambda-3\beta \epsilon+\Or(\epsilon^2)] (\tilde{W}_0+R \tilde{W}_1+\ldots) \nonumber \\
=Re_{\rm p} \left(-1+\beta^3\right)\frac{1}{2}\lambda\frac{\rmd^2 \lambda}{\rmd \tau^2} +
\frac{9}{4}\left[-1+\beta + \Or(\epsilon^2)\right]\lambda \frac{\rmd \lambda}{\rmd \tau} \label{eq3.4}.
\end{eqnarray}
The solution to $\Or(Re_{\rm p}^0)$ is
\begin{equation}
\tilde{W}_0 = \frac{1}{2}\frac{(-1+\beta)\lambda
\rmd \lambda / \rmd \tau}{(1+\beta)\lambda-3\beta\epsilon} + \Or(\epsilon^2) \label{eq3.5},
\end{equation}
and $\langle \hat{W}_0 \rangle = 0$. Note that the average translational velocity would remain zero even if higher-order terms were retained, i.e., there is no net
translation of the two-sphere system to $\Or(Re_{\rm p}^0)$. In contrast, the solution to $\Or(Re_{\rm p}^1)$ includes terms of the form $\lambda \; (\rmd^2 \lambda/\rmd \tau^2)$, which yield a non-zero average. For example, suppose the distance between the spheres varies as $\lambda = 1+\epsilon \tilde{A}\sin\tau$, where $\tilde{A} = A / a =\Or(1)$ is the non-dimensional amplitude. In this case,
\begin{equation}
\langle \tilde{W}_1 \rangle =
\epsilon^3\frac{\beta(\beta-1)\left(-\beta^2+3\beta-1\right)}{6(1+\beta)^2}
\tilde{A}^2 + \Or(\epsilon^4) \label{eq3.7}.
\end{equation}
Then, in dimensional form, the net translational velocity is
\begin{equation}
\langle W \rangle = Re_{\rm p} a \omega \epsilon^2
\frac{\beta(\beta-1)\left(-\beta^2+3\beta-1\right)}{6(1+\beta)^2}
(A/a)^2 + \Or(Re_{\rm p} \epsilon^3,Re_{\rm p}^2) \label{eq3.8}.
\end{equation}
Note that $\langle W \rangle$ tends to zero as $L\rightarrow \infty$,
reflecting the fact that the net translational velocity is due
to the interaction between the spheres. The smaller sphere advances in front
($\langle W \rangle > 0$) for $1<\beta<(3+\sqrt{5})/2$, and the
larger sphere advances in front ($\langle W \rangle < 0$) for $\beta>(3+\sqrt{5})/2$.

\subsection{Circular motion of two unequal spheres}

\begin{figure} [htbp]
\begin{center}
\includegraphics[width=3in]{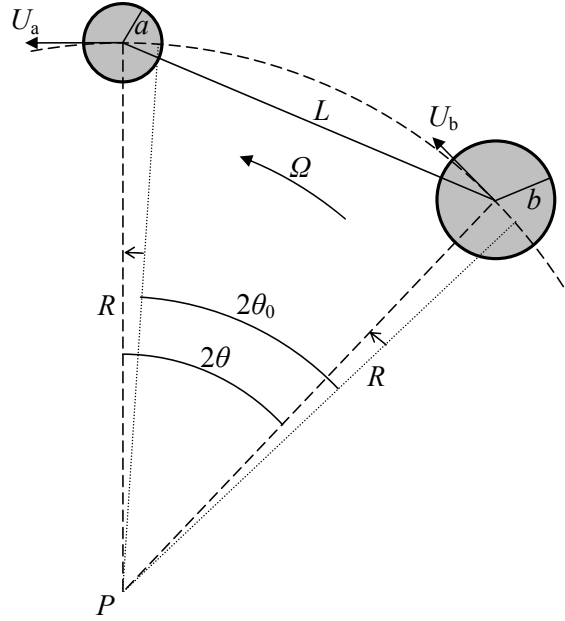}
\end{center}
\caption{The inertial rotator. The spheres are constraint to move along the circumference of centre $P$ and radius $R$.} \label{fig1b}
\end{figure}

Consider the swimmer shown in figure \ref{fig1b}, which is a two-sphere
version of the rotator devised by Dreyfus \etal
\cite{DreyfusBaudryEtAl05}. As was the case in the previous example, the spheres have radii $a$ and $b$,
which are different but of comparable magnitude, that is, $\beta \equiv b/a
= \Or(1)$. Again, $\beta\neq 1$ is required to break the spatial symmetry and yield net motion. The spheres are constrained to move along the circumference of centre
$P$ (which is held fixed) and radius $R$, which is much larger than
$a$, $\hat R \equiv R/a \gg 1$. The angle between the spheres, $2\theta(t)$, varies periodically with
radian frequency $\omega$ in a prescribed manner. Specifically, $\theta(t)=\theta_0 + \epsilon \theta_1(t)$, with $\epsilon \equiv a/\langle L \rangle$, $L$ is the distance between the sphere centres, and $\Or(\theta_0)=\Or(\theta_1)=\Or(1)$. As a result of the interaction between the spheres, there is an unknown
rigid-body rotation with angular velocity $\Omega(t)$, whose time average we want to determine. The spheres
move with linear velocities $U_{\rm a}(t)$ and $U_{\rm b}(t)$ and experience
hydrodynamic resistance forces $F^{\rm h}_{\rm a}(t)$ and $F^{\rm h}_{\rm b}(t)$ in the azimuthal direction.

If both spheres were towed along the circumference at a constant angular velocity $\hat{\Omega}$, the azimuthal hydrodynamic forces on the spheres, taken positive in counterclockwise direction, would be \cite{HappelBrenner65}
\numparts
\begin{eqnarray}
\hat F^{\rm h}_{\rm a} &=& 6 \pi \mu a R \hat{\Omega} \left[-1 + \frac{3 b}{4
R}\frac{\cos^2\theta}{\sin \theta}-\frac{3 b}{8 R}\sin
\theta+\Or(\epsilon^3)\right] \label{eq4.1} \\
\hat F^{\rm h}_{\rm b} &=& 6 \pi \mu b R \hat{\Omega} \left[-1 + \frac{3 a}{4
R}\frac{\cos^2\theta}{\sin \theta}-\frac{3 a}{8 R}\sin
\theta+\Or(\epsilon^3)\right]. \label{eq4.2}
\end{eqnarray}
\endnumparts
Introducing the expression of the prescribed motion into (\ref{eq4.1}) and (\ref{eq4.2}) yields
\numparts
\begin{eqnarray}
\hat F^{\rm h}_{\rm a} &=& 6 \pi \mu a R \hat{\Omega} \left[-1 + \frac{3 b}{8R}\left(\gamma_0-\gamma_1\epsilon \theta_1+\Or(\epsilon^2)\right)\right] \label{eq4.3}\\
\hat F^{\rm h}_{\rm b} &=& 6 \pi \mu b R \hat{\Omega} \left[-1 + \frac{3 a}{8R}\left(\gamma_0-\gamma_1\epsilon \theta_1+\Or(\epsilon^2)\right)\right], \label{eq4.4}
\end{eqnarray}
\endnumparts
where $\gamma_0 = (3\cos^2 \theta_0 -1)/\sin \theta_0$ and $\gamma_1
= \cos \theta_0 (2+3\sin^2 \theta_0)/\sin^2 \theta_0$. With (\ref{eq4.3}) and (\ref{eq4.4}), application of (\ref{eq2.3}) results in
\begin{eqnarray}
\fl \rho_{\rm p} \frac{4}{3}\pi \left[(a^3+b^3)R \frac{\rmd \Omega}{\rmd t}+(a^3-b^3)R\frac{\rmd^2 \theta}{\rmd t^2}\right]R\hat{\Omega} + \nonumber \\ \rho_{\rm p}\frac{4}{3}\pi \left[\frac{2}{5}(a^5+b^5) + \frac{a^9+b^9}{\left(a^3+b^3\right)^2}L^2 \right]\frac{\rmd \Omega}{\rmd t}\hat{\Omega} - (\hat F^{\rm h}_{\rm a}R+\hat F^{\rm h}_{\rm b}R)\Omega \nonumber \\ = (\hat F^{\rm h}_{\rm a}-\hat F^{\rm h}_{\rm b}) R \frac{\rmd \theta}{\rmd t}. \label{eq4.5}
\end{eqnarray}
Next, we non-dimensionalize (\ref{eq4.5}) by defining $Re_{\rm p} \equiv \rho_{\rm p}
a^2 \omega / \mu \ll \epsilon \ll 1$, $\tau \equiv \omega t$, $\beta \equiv b/a$, $\tilde
R \equiv R \epsilon /a$, and $\tilde \Omega \equiv \Omega / (\epsilon \omega) $ and expand $\tilde \Omega = \tilde \Omega_0 + Re_{\rm p} \tilde \Omega_1
+\ldots$. Thus,
\begin{eqnarray}
\fl \frac{2}{9}Re_{\rm p} \tilde R \left[\left(\beta^3+1\right)+\frac{\beta^9+1}{\left(\beta^3+1\right)^2}\sin^2 \theta_0 + \Or(\epsilon)\right] \left(\frac{\rmd \tilde \Omega_0}{\rmd \tau}
+\ldots\right) \nonumber \\ +
\left[(1+\beta)\tilde R -
\frac{3}{4}\beta\left(\gamma_0 \epsilon-\gamma_1\epsilon^2\theta_1+\Or(\epsilon^3)\right)\right]
\left(\tilde \Omega_0 + Re_{\rm p} \tilde \Omega_1
+\ldots\right) \nonumber \\
= \frac{2}{9}Re_{\rm p} \tilde R \left(\beta^3-1\right) \frac{\rmd^2 \theta_1}{\rmd \tau^2}
+(\beta-1)\tilde R \frac{\rmd \theta_1}{\rmd \tau}.
\label{eq4.6}
\end{eqnarray}
The solution to $\Or(Re_{\rm p}^0)$ is
\begin{eqnarray}
\fl \tilde \Omega_0 = \frac{(\beta-1)}{(\beta+1)}\frac{\rmd \theta_1}{\rmd \tau}\left[1+\frac{3}{4}\frac{\beta}{(1+\beta)}\frac{\gamma_0}{\tilde R}\epsilon \nonumber \right. \\ \left. +\left(\frac{9}{16}\frac{\beta^2}{(1+\beta)^2}\frac{\gamma_0^2}{\tilde R^2}-\frac{3}{4}\frac{\beta}{(1+\beta)}\frac{\gamma_1 \theta_1}{\tilde R}\right)\epsilon^2+\Or(\epsilon^3)\right].
\label{eq4.7}
\end{eqnarray}
Since $\theta_1$ is periodic, $\langle \tilde \Omega_0\rangle=0$.
However the solution to $\Or(Re_{\rm p}^1)$ includes terms of the form
$\theta_1 \rmd^2 \theta_1 / \rmd \tau^2$, which yield a net angular velocity. If we
consider the case $\theta_1 = \tilde A/(2\tilde R) \sin \tau$, where $\tilde A =A/a$ is the non-dimensional amplitude of motion, the
time-averaged non-dimensional angular velocity becomes
\begin{eqnarray}
\fl \langle \tilde \Omega_1 \rangle= \epsilon^2 \frac{\gamma_1}{48 \tilde R^3}\frac{\beta (\beta-1)}{(\beta+1)^2}\left[\left(-\beta^2+3\beta-1\right)-\frac{2\left(\beta^9+1\right)}{\left(\beta^3+1\right)^3}\sin^2\theta_0\right]\tilde A^2 \nonumber \\+ \Or(\epsilon^3). \label{eq4.8}
\end{eqnarray}
It is noted that this non-dimensional velocity becomes zero when $\theta_0=\pi/2$, as this corresponds to the fore-aft symmetric configuration in which the spheres are diametrally opposed. The velocity becomes unboundedly large if $\tilde R$ remains constant and $\theta_0 \rightarrow 0$. In this limit, however, the spheres become close and (\ref{eq4.8}) is invalid. According to (\ref{eq4.8}), the smaller sphere advances in front for $1<\beta<1.74$, while the larger sphere advances in front for $\beta>(3+\sqrt{5})/2 \approx 2.62$. For $1.74<\beta<2.62$, the direction of rotation depends on the value of $\theta_0$.

In dimensional form, the net angular velocity is
\begin{equation}
\langle \Omega \rangle= \omega \epsilon Re_{\rm p} \langle \tilde \Omega_1
\rangle + \Or(\epsilon^4 Re_{\rm p}, Re_{\rm p}^2). \label{eq4.9}
\end{equation}
 Note that the rotator reduces to the previous example of two unequal spheres moving along their line of centres for $\tilde R \rightarrow \infty$ with constant $\epsilon = 1/(2 \tilde R \sin \theta_0)$. In this limit, $\sin \theta_0 \rightarrow 0$, $\gamma_1 \rightarrow 8 \tilde R^2$, and the net linear velocities predicted by (\ref{eq3.8}) and (\ref{eq4.9}) agree.

\subsection{Translation of a scallop-like swimmer}

\begin{figure} [htbp]
\begin{center}
\includegraphics[width=3in]{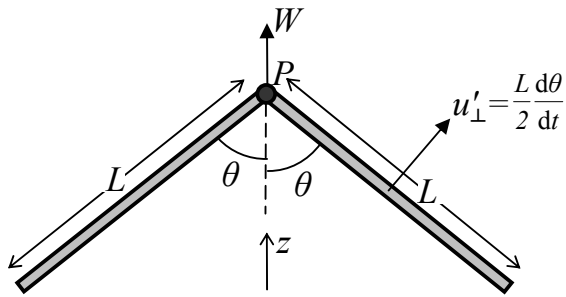}
\end{center}
\caption{Scallop-like swimmer consisting of two circular cylinders of radius $a$ and length $L$ connected by a hinge at $P$.} \label{fig2}
\end{figure}

We study the scallop-like swimmer schematized in figure \ref{fig2}.
The swimmer consists of two circular cylinders of radius $a$ and
length $L \gg a$ connected by a hinge. The angle between the two cylinders, $2 \theta
(t)$, is a prescribed periodic function of time with radian
frequency $\omega$. The swimmer undergoes an unknown translational
velocity, $W(t)$, whose time average we want to determine.

\begin{figure} [htbp]
\begin{center}
\includegraphics[width=3in]{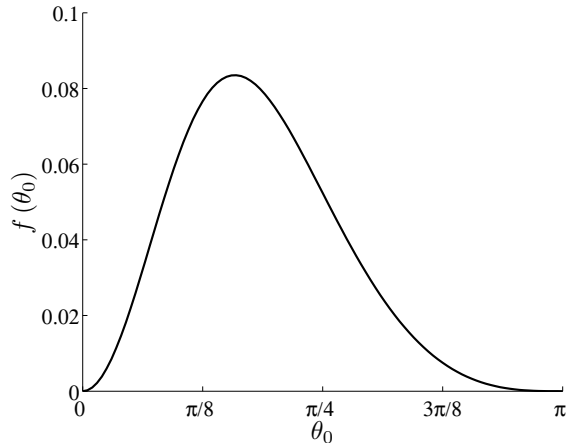}
\end{center}
\caption{Dependency of the net translational velocity on the average angle of opening of the scallop-like swimmer, $\theta_0$.} \label{fig2b}
\end{figure}

By neglecting hydrodynamic interactions between the cylinders, the hydrodynamic forces exerted on each cylinder are related to the velocities of
the cylinder by the resistance coefficients. The resistance coefficients for translation perpendicular and parallel to the cylinder axis are
$R_\perp = 2R_\parallel=4\pi\mu L / \ln(L/a)$
\cite[e.g.,][]{Wegener04}. Application of (\ref{eq2.3}) in the $z$-direction yields
\begin{eqnarray}
\fl m \left[\frac{\rmd W}{\rmd t}+\frac{\rmd}{\rmd t}\left(\frac{L}{2}\frac{\rmd \theta}{\rmd t} \sin \theta\right)\right]\hat W - \left[\left(-R_\parallel \hat W \cos \theta\right)W \cos \theta \right. \nonumber \\ \left. + \left(-R_\perp \hat W \sin \theta\right) W \sin \theta \right] = \left(-R_\perp \hat W \sin \theta\right)\frac{L}{2}\frac{\rmd \theta}{\rmd t} \label{eq5.2},
\end{eqnarray}
where $m=\rho_{\rm p} \pi a^2 L$ is the mass
of each cylinder. This equation is made non-dimensional by using $a$ and $1/\omega$ as
length and time scales and by defining a non-dimensional time $\tau
\equiv \omega t$, cylinder length $\lambda \equiv L/a$, and velocity
$\tilde{W} \equiv W/(\omega L)$. Introducing these
scalings into (\ref{eq5.2}) and expanding $\tilde{W}$ in powers of $Re_{\rm p} \equiv \rho_{\rm p} a^2 \omega / \mu \ll \epsilon \ll 1$ result in
\begin{eqnarray}
\fl Re_{\rm p} \left( \frac{\rmd \tilde{W}_0}{\rmd \tau} + \ldots \right) + \frac{2}{\ln
\lambda}\left(1+\sin^2 \theta\right) \left(\tilde{W}_0 + Re_{\rm p}\tilde{W}_1 +\ldots \right) \nonumber \\
= -\frac{Re_{\rm p}}{2}\frac{\rmd}{\rmd \tau}\left(\sin \theta \frac{\rmd \theta}{\rmd \tau}\right) - \frac{2}{\ln \lambda}\sin
\theta \frac{\rmd \theta}{\rmd \tau} \label{eq5.3}.
\end{eqnarray}
The solution to $\Or(Re_{\rm p}^0)$ is
\begin{equation}
\tilde{W}_0 = -\frac{\sin \theta}{1+\sin^2 \theta}\frac{\rmd \theta}{\rmd \tau} \label{eq5.4}
\end{equation}
and $\langle \tilde{W}_0 \rangle = 0$. Let us assume small oscillations of the form $\theta (\tau) =
\theta_0 + \theta_1 (\tau) = \theta_0 + \epsilon \sin \tau$, where
$\epsilon \ll 1$. Then, the solution to $\Or(Re_{\rm p}^1)$ has a non-zero average given by
\begin{equation}
\langle \tilde{W}_1 \rangle = \frac{1}{2}\epsilon^2 \ln(\lambda)
f(\theta_0) + \Or(\epsilon^2) \label{eq5.5},
\end{equation}
where
\begin{equation}
f(\theta_0) = \frac{\sin^2 \theta_0 \cos^3 \theta_0}{\left(1+\sin^2\theta_0\right)^3}
\label{eq5.6}.
\end{equation}
The dimensional net
velocity is $\langle W \rangle = Re_{\rm p} \omega L \langle \tilde{W}_1
\rangle + \Or(Re_{\rm p}^2)$. The function $f(\theta_0)$ is represented in figure \ref{fig2b}. Since $f(\theta_0)>0$ for all $\theta_0 \in
(0,\pi/2)$, $\langle \tilde{W}_1 \rangle > 0 $, and the swimmer moves
with the hinge advancing in front. The net velocity is maximum for $\theta_0 = \arccos([-7/2+\sqrt{73}/2]^{1/2}) \approx 28.5^\circ$.

\subsection{Translation of a deforming sphere}

\begin{figure} [htbp]
\begin{center}
\includegraphics[width=2in]{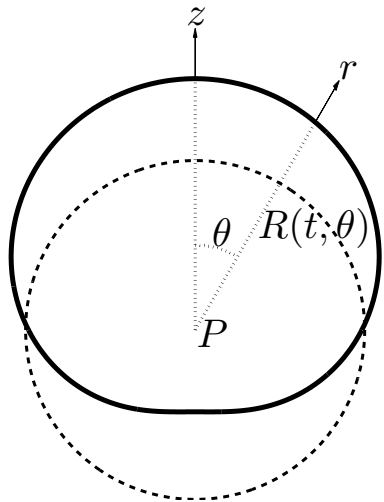}
\end{center}
\caption{Undeformed (dashed line) and deformed (solid line) geometry of the swimmer. The deformation is axisymmetric with respect to the $z$-axis. Not to scale.}\label{fig4}
\end{figure}

A quasi-spherical swimmer of radius $a$ executes a small, periodic deformation, so that its radius is
given by $R(t,\theta)=a(1+\epsilon^2 \alpha_0(t) P_0(\cos \theta)+\epsilon \alpha_1(t) P_1(\cos \theta))$, as shown in figure \ref{fig4}. Here, $P_0 \equiv 1$ and $P_1 \equiv \cos \theta$ are the
Legendre polynomials of orders $0$ and $1$, respectively, and $\epsilon \ll 1$. The coefficient of $P_0$ is chosen as $\alpha_0=- \alpha_1 /3 + \Or(\epsilon^4)$, so that the particle volume remains constant and equal to $4 \pi a^3/3$. The deformation is axisymmetric
with respect to the $z$-axis, and it is such that the particle density, $\rho_p$, remains homogeneous and constant.

To compute the translation of the sphere, we apply the $z$-component of (\ref{eq2.3}) to this swimmer, defining the reference point $P$ as the centre of the undeformed sphere. To evaluate the first term on the left-hand side of (\ref{eq2.3}), we write the velocity of the centre of mass, $G$, as $u_{\rm G}=u_{\rm P}+u'_{\rm G}$, where $u'_{\rm G}$ is the velocity of $G$ relative to $P$. The acceleration of $G$ relative to $P$ is
\begin{equation}
\frac{\rmd u'_{\rm G}}{\rmd t} =  a
 \frac{\rmd ^2 \alpha_1}{\rmd t^2} \left[\frac{4}{5} \epsilon + \Or\left(\epsilon^3\right)\right], \label{eq6.1}
\end{equation}
directed along the $z$-axis. The normal vector to the deformed sphere's surface in spherical
coordinates $(r,\theta,\phi)$ is
\begin{equation}
\boldsymbol{n} = \left(\begin{array}{c}
                         \frac{1}{\left[1+(\rmd R/\rmd \theta)^2/R^2\right]^{1/2}} \\
                         \frac{-(\rmd R/\rmd \theta)/R}{\left[1+(\rmd R/\rmd \theta)^2/R^2\right]^{1/2}} \\
                         0
                       \end{array}\right) =
                 \left(\begin{array}{c}
                         1-\epsilon^2 \alpha_1^2 V_1^2/2  \\
                         \epsilon \alpha_1V_1-\epsilon^2\alpha_1^2P_1V_1 \\
                         0
                       \end{array}\right) + \Or(\epsilon^2),
                       \label{eq6.2}
\end{equation}
where we have used Lighthill's \cite{Lighthill52} definition,
\begin{equation}
V_n(\eta) \equiv \frac{2\sqrt{1-\eta^2}}{n(n+1)}\frac{\rmd P_n(\eta)}{\rmd \eta},
\label{eq6.3}
\end{equation}
with $\eta \equiv \cos \theta$. To evaluate the two surface integrals in (\ref{eq2.3}), we must calculate the flow around the
rigid, deformed sphere past a uniform flow, $\boldsymbol{\hat U}$.
Any axisymmetric flow can be described by the streamfunction
\begin{equation}
\psi = \sum_{n=0}^\infty \left[A_n r^{n+3} + B_n r^{n+1} + C_n
r^{2-n} +D_n r^{-n}\right] Q_n(\eta), \label{eq6.4}
\end{equation}
where $A_n$, $B_n$, $C_n$, and $D_n$ are constants, and
$\rmd Q_n(\eta) / \rmd \eta=P_n(\eta)$ \cite[e.g.,][]{Leal92}. The radial and
azimuthal velocity components, $u_r$ and $u_\theta$, are given by
\numparts
\begin{eqnarray}
u_r &=& -\frac{1}{r^2}\frac{\partial \psi}{\partial \eta} \label{eq6.5a} \\
u_\theta &=& -\frac{1}{r\sqrt{1-\eta^2}}\frac{\partial \psi}{\partial r}.
\label{eq6.5b}
\end{eqnarray}
\endnumparts
In our problem, in which the reference system is attached to the swimmer, $u_r$ and $u_\theta$ must tend to the ambient velocity, of magnitude $\hat U$, as
$r \rightarrow \infty$. In addition, the solutions
with $n=0$ are unadmissible, because the corresponding azimuthal
velocities are undefined at $\eta=1$ ($\theta=0$). With these
constraints, the general solution of the flow field is
\numparts
\begin{eqnarray}
u_r &=& -\hat U \cos \theta + \sum_{n=1}^\infty \left(A_n
\frac{a^n}{r^n}+B_n\frac{a^{n+2}}{r^{n+2}}\right)P_n \label{eq6.6a} \\
u_\theta &=& \hat U \sin \theta + \sum_{n=1}^\infty \left(A_n
\left(\frac{n}{2}-1\right)\frac{a^n}{r^n}+B_n\frac{n}{2}\frac{a^{n+2}}{r^{n+2}}\right)V_n,
\label{eq6.6b}
\end{eqnarray}
\endnumparts
as obtained in previous studies of spherical swimmers \cite{Lighthill52,Blake71}. The constants are determined using the boundary conditions, $u_r=u_\theta=0$,
at $r=R=a(1+\epsilon^2 \alpha_0 P_0 + \epsilon\alpha_1P_1)$. These constants can be expanded in powers of $\epsilon$:
\numparts
\begin{eqnarray}
A_n &=& \sum_{k=0}^\infty \epsilon^k A_n^{(k)} \label{eq6.7a} \\
B_n &=& \sum_{k=0}^\infty \epsilon^k B_n^{(k)}. \label{eq6.7b}
\end{eqnarray}
\endnumparts
The solution is
\numparts
\begin{eqnarray}
A_1 &=& \hat{U}\left[\frac{3}{2}
+\epsilon^2\frac{3}{2}\alpha_0 +\Or(\epsilon^3)\right] \label{eq6.8a} \\
B_1 &=& \hat{U}\left[-\frac{1}{2}
+\epsilon^2\left(-\frac{3}{2}\alpha_0+\frac{9}{10}\alpha_1^2\right) +\Or(\epsilon^3)\right]\label{eq6.8b} \\
A_3 &=& \hat{U}\left[-\epsilon^2\frac{9}{10}\alpha_1^2 +\Or(\epsilon^3)\right]\label{eq6.8c} \\
B_3 &=& \hat{U}\left[\epsilon^2\frac{3}{2}\alpha_1^2
+\Or(\epsilon^3)\right], \label{eq6.8d}
\end{eqnarray}
\endnumparts
while all other constants are at most of $\Or(\epsilon^3)$. The value of the pressure on the deformed sphere's surface
is
\begin{eqnarray}
\fl p(R) = \frac{\mu\hat
U}{a}\left[\frac{3}{2}P_1-3\epsilon\alpha_1P_1^2 +
\epsilon^2\left(-\frac{3}{2}\alpha_0 P_1 +\frac{9}{2}\alpha_1^2 P_1^3-\frac{9}{4}\alpha_1^2 P_3\right)\right. \nonumber \\ \left.+\Or(\epsilon^3)\right].
\label{eq6.9}
\end{eqnarray}
The radial and tangential stresses at $r=R$ are
\numparts
\begin{eqnarray}
\fl \left. \sigma_{rr} \right|_{r=R} = \frac{\mu \hat
U}{a}\left[-\frac{3}{2}P_1-\epsilon\alpha_1 3
P_1^2\right. \nonumber \\
\left.+\epsilon^2\left(\frac{33}{2}P_1^3+\left(\frac{3}{2}\alpha_0-\frac{27}{5}\alpha_1^2\right)P_1+\frac{237}{20}\alpha_1^2 P_3\right)+\Or(\epsilon^3)\right]
\label{eq6.10} \\
\fl \left. \sigma_{r\theta} \right|_{r=R} = \frac{\mu \hat
U}{a}\left[\frac{3}{2}V_1-\epsilon\alpha_1 6 P_1V_1\right. \nonumber \\\
 \left. +\epsilon^2 \left(\frac{27}{2}\alpha_1^2 P_1^2V_1+
\left(-\frac{3}{2}\alpha_0-\frac{27}{10}\alpha_1^2\right)V_1-\frac{153}{10}\alpha_1^2 V_3\right)\right. \nonumber \\ \left.+\Or(\epsilon^3)\right]
\label{eq6.11} \\
\fl \left. \sigma_{\theta \theta} \right|_{r=R} = \frac{\mu \hat
U}{a}\left[-\frac{3}{2}P_1+\epsilon\alpha_1 6
P_1^2+\Or(\epsilon^2)\right] \label{eq6.12}
\end{eqnarray}
\endnumparts
Finally, the two surface integrals in (\ref{eq2.3}) are
\begin{eqnarray}
\fl \hat F_z = \int_{S(t)}\boldsymbol{n} \cdot \boldsymbol{\hat \sigma} \cdot
\boldsymbol{\hat z}\; \rmd S = \int_0^{\pi} \boldsymbol{n} \cdot
\left(\begin{array}{c}
                                                                   \cos \theta \sigma_{rr} - \sin \theta \sigma_{r\theta} \\
                                                                   \cos \theta \sigma_{r\theta} - \sin \theta \sigma_{\theta \theta} \\
                                                                   \cos \theta \sigma_{r\phi} - \sin \theta \sigma_{\theta \phi}
                                                                 \end{array}
\right) \left(2\pi R^2\sin\theta \rmd \theta\right) \nonumber \\ = -6
\pi \mu \hat U
a\left[1+\epsilon^2\left(\alpha_0-\frac{7}{15}\alpha_1^2\right)+\Or(\epsilon^3) \right]
\label{eq6.13} \\
\fl \int_{S(t)}\boldsymbol{n} \cdot \boldsymbol{\hat \sigma} \cdot
\boldsymbol{u'} \; \rmd S = \int_0^{\pi}\boldsymbol{n} \cdot
\left(\begin{array}{c}
        \sigma_{rr} \\
        \sigma_{r\theta} \\
        \sigma_{r\phi}
      \end{array}
\right)a \left(\epsilon^2 \frac{\rmd \alpha_0}{\rmd t} + \epsilon P_1 \frac{\rmd \alpha_1}{\rmd t}\right)\left(2\pi
R^2\sin\theta \rmd \theta\right) \nonumber \\
= \pi \mu \hat U a^2
\left[-2\epsilon\frac{\rmd \alpha_1}{\rmd t}+\epsilon^3\left(-2\alpha_1^2\frac{\rmd \alpha_1}{\rmd t}-2\alpha_0\frac{\rmd \alpha_1}{\rmd t}-4\alpha_1\frac{\rmd \alpha_0}{\rmd t}\right)+\Or(\epsilon^4)\right].
\label{eq6.14}
\end{eqnarray}
Introducing these results into (\ref{eq2.3}) yields, in non-dimensional form,
\begin{eqnarray}
\fl \frac{4}{3}Re_{\rm p} \frac{\rmd \tilde U}{\rmd \tau} +
6\left[1+\epsilon^2\left(-\frac{1}{3}\alpha_1-\frac{7}{15}\alpha_1^2\right)+\Or(\epsilon^3)\right]\tilde
U \nonumber \\
=
-2\frac{\rmd \alpha_1}{\rmd \tau}+\epsilon^2\left(-2\alpha_1^2 +2\alpha_1\right)\frac{\rmd \alpha_1}{\rmd t}-\frac{16}{15}Re_{\rm p}\frac{\rmd^2\alpha_1}{\rmd \tau^2}+\Or(\epsilon^3),
\label{eq6.15}
\end{eqnarray}
where $\tau \equiv \omega t$, $U \equiv \epsilon \omega
a \tilde{U}$.
Next, we expand $\tilde U = \tilde U_0 + Re_{\rm p} \tilde U_1 + \ldots$ and assume
$Re_{\rm p} \ll \epsilon \ll 1$. The solution to $\Or(Re_{\rm p}^0)$ is
\begin{equation}
\tilde U_0 =
-\frac{1}{3}\frac{\rmd \alpha_1}{\rmd \tau}+\epsilon^2\left(-\frac{22}{45}\alpha_1^2+\frac{2}{9}\alpha_1\right)\frac{\rmd \alpha_1}{\rmd \tau}+\Or(\epsilon^3).
\label{eq6.16}
\end{equation}
For any periodic $\alpha_1(\tau)$, the time average of this
expression is zero. Moreover, the time average is zero to all orders of
$\epsilon$.

The solution to $\Or(Re_{\rm p}^1)$ includes terms of the form $\epsilon^2
(\rmd \alpha_1/\rmd\tau)^2$ and $\epsilon^2\alpha_1
(\rmd \alpha_1/\rmd\tau)^2$ which, for a suitable choice of $\alpha_1(t)$, yield a non-zero average. For instance, for $\alpha_1 = \sin^2
(\omega t) = \sin^2 \tau$, the time-averaged non-dimensional
velocity is
\begin{equation}
\langle \tilde U_1 \rangle =
-\frac{2}{405}\epsilon^2+\Or(\epsilon^3),
\label{eq6.17}
\end{equation}
or, in dimensional form,
\begin{equation}
\langle U \rangle = -
\frac{2}{405}Re_{\rm p} \epsilon^3 \omega a +
\Or(\epsilon^4 Re_{\rm p}, Re_{\rm p}^2). \label{eq6.18}
\end{equation}
Note that the minus sign indicates a net translation in the $(-z)$-direction.

\section{Conclusion\label{sec4}}
We have presented several examples of dense reciprocal swimmers that are able to propel themselves, even in the absence of fluid inertia, at arbitrarily small values of $Re_{\rm p}$. The existence of a net swimming velocity arises from a nonlinear interaction between the oscillatory particle inertia and the oscillatory drag in the presence of spatial asymmetries. The existence of this nonlinear interaction can be inferred from the reciprocal theorem, (\ref{eq2.3}). In the absence of rigid-body rotation, (\ref{eq2.3}) reads
\begin{equation}
\left(m \frac{\rmd \boldsymbol{u_{\rm G}} }{\rmd t}\right) \cdot \boldsymbol{\hat{U}_{\rm P}} - \boldsymbol{\hat F} \cdot \boldsymbol{U_{\rm P}} = \int_{S(t)} \bi{n} \cdot \boldsymbol{\hat{\sigma}} \cdot \boldsymbol{u'} \; \rmd S. \label{eq7.0.5}
\end{equation}
$\boldsymbol{\hat F}$, the hydrodynamic drag on the towed, deformed body, scales as $(\mu a \hat U_{\rm P}) (1+A/a \, \alpha(t))$. Here, $(\mu a \hat U_{\rm P})$ is the magnitude of the hydrodynamic drag on the towed, undeformed body, $A$ is the amplitude of oscillation of the swimmer surface, $\alpha(t)$ is the non-dimensional oscillation,  and $a \gg A$ is the size of the undeformed swimmer. The integral on the right-hand side is typically of magnitude $(\mu a \hat U_{\rm P}) (K_1 + A/a \, K_2 \alpha(t)) u'$, where $u'=A \; \rmd \alpha(t) / \rmd t$ is the magnitude of the deformational surface velocity $\boldsymbol{u'}$, and $K_1$ and $K_2$ are coefficients that account for the change of magnitude and direction of $\boldsymbol{u'}$ over the swimmer surface. Also, $u_{\rm G} \sim U_{\rm P} \sim A \omega$. Introducing these scalings into (\ref{eq7.0.5}) yields
\begin{equation}
Re_p \frac{\rmd \tilde U_{\rm P}}{\rmd \tau} - \left(1+\epsilon\alpha(\tau)\right) \tilde U_{\rm P}  \sim \left(K_1 + \epsilon K_2 \alpha(\tau)\right) \frac{\rmd \alpha}{\rmd \tau},  \label{eq7.1}
\end{equation}
where $\tilde U_{\rm P}$ is the non-dimensional translational velocity, $\tau \equiv \omega t$ is the non-dimensional time, and $\epsilon \equiv a/A \ll 1$. By expanding $\tilde U_{\rm P}$ in powers of $Re_{\rm p}$, the equation to $\Or(Re_{\rm p}^0)$ yields
\begin{equation}
\tilde U_{{\rm P}, 0}  \sim - \left[K_1 + \epsilon \left(K_2-K_1\right) \alpha(\tau) + \Or(\epsilon^2)\right]\frac{\rmd \alpha}{\rmd \tau},
\end{equation}
and $\langle \tilde U_{{\rm P},0} \rangle = 0$, in agreement with the scallop theorem. Next, the equation to $\Or(Re_{\rm p}^1)$ yields
\begin{equation}
\tilde U_{{\rm P}, 1} \sim \frac{\rmd \tilde U_{{\rm P},0}}{\rmd \tau} \left[1 - \epsilon\alpha(\tau) + \Or(\epsilon^2)\right]. \label{eq7.2.6}
\end{equation}
 Thus, in the presence of certain spatial asymmetries, the right-hand side of (\ref{eq7.2.6}) contains a term of the form
\begin{equation}
\epsilon \left[ \left(K_2-2K_1\right) \alpha(\tau) \frac{\rmd^2 \alpha}{\rmd \tau^2} + \left(K_2-K_1\right) \left(\frac{\rmd \alpha}{\rmd \tau}\right)^2  \right].\label{eq7.3}
\end{equation}
For an appropriate choice of the reciprocal oscillation $\alpha(\tau)$ (for example, a sinusoidal function), this term gives raise to a non-zero net velocity to $\Or(Re_{\rm p}^1)$. Accordingly, in the four examples presented in this paper, the net translational velocity is proportional to $Re_{\rm p}$. As discussed in section \ref{sec2}, a necessary condition for the existence of this non-zero velocity is that the dense swimmer undergoes deformation normal to its surface, while pure squirming does not result in locomotion.

In the previous analysis we have not included the effect of gravity, which needs to be considered in practical applications. For $A \sim a$ and $\epsilon=\Or(1)$, we expect the net translational velocity due to the reciprocal motion to scale as
\begin{equation}
\langle U_p \rangle \sim a \omega Re_{\rm p}, \label{eq7.4}
\end{equation}
while a typical sedimentation speed is
\begin{equation}
U_g \sim \frac{\rho_p a^2 g}{\mu}. \label{eq7.5}
\end{equation}
Thus, the dense swimmer needs to oscillate at a frequency $\omega > O(\sqrt{g/a})$ in order to overcome gravity.

In conclusion, our examples show the existence of a net translation or rotation at any arbitrarily small $Re_{\rm p} >0$ for certain spatially asymmetric, dense reciprocal swimmers. This demonstrates the breakdown of the scallop theorem for a case in which the flow around the swimmer is governed by the Stokes equations, that is, in the absence of fluid inertia. While we have studied the limit of dense swimmers, for which $\rho_p \gg \rho$, in many biological applications $\rho_p \sim \rho$ and the unsteady fluid inertia is expected to be comparable to the particle inertia. The extension of this study to the limit of $Re_{\rm f} \ll Re_{\rm \omega} \sim Re_{\rm p}$ will be the subject of future work.

\bibliographystyle{unsrt}

\end{document}